# Large Language Model Agents Enable Autonomous Design and Image Analysis of Microwell Microfluidics

Dinh-Nguyen Nguyen[1], Sadia Shakil[1], Raymond Kai-Yu Tong[1*], Ngoc-Duy Dinh[1*]

[1]Department of Biomedical Engineering, The Chinese University of Hong Kong, Shatin, N.T., Hong Kong, China

*Corresponding Author

Corresponding Email: ngocduydinh@cuhk.edu.hk

## Abstract

Microwell microfluidics has been utilized for single-cell analysis to reveal heterogeneity in gene expression, signaling pathways, and phenotypic responses for identifying rare cell types, understanding disease progression, and developing more precise therapeutic strategies. However, designing microwell microfluidics is a considerably complex task, requiring knowledge, experience, and CAD software, as well as manual intervention, which often fails initial designs, demanding multiple costly and time-consuming iterations. In this study, we establish an autonomous large language model (LLM)-driven microwell design framework to generate code-based computer-aided design (CAD) scripts, that enables the rapid and reproducible creation of microwells with diverse geometries and imaging-based analysis. We propose a multimodal large language model (MLLM)–logistic regression framework based on integrating high-level semantic descriptions generated by MLLMs with image embeddings for image classification tasks, aiming to identify microwell occupancy and microwell shape. The fused multimodal representation is input to a logistic regression model, which is both interpretable and computationally efficient. We achieved significant improvements, exceeding 0.92 for occupancy classification and 0.99 for shape classification, across all evaluated MLLMs, compared with 0.50 and 0.55, respectively, when relying solely on direct classification. The MLLM–logistic regression framework is a scalable, efficient solution for high-throughput microwell image analysis. Our study demonstrates an autonomous design microwell platform by translating natural language prompts into optimized device geometries, CAD scripts and image analysis, facilitating the development of next-generation digital discovery by integration of literature mining, autonomous design and experimental data analysis.

KEYWORDS: Microwell Microfluidics, Large Language Model Agents, Large Language Models, Multimodal Large Language Models, Autonomous Design, Artificial Intelligence

1. Introduction

Microfluidics has provided an essential tool for precisely manipulating and controlling small amounts of fluids at the microscale, offering lower reagent consumption, faster analysis of biochemical reactions, high throughput screening, and automation[1,2]. Microfluidics includes channel microfluidics[3–5], droplet microfluidics[6–12], and microwell microfluidics. In three types of microfluidic platforms, microwell-microfluidics has demonstrated the advantages such as simple operation and control, scalability and compatibility with AI-driven analysis. Microwell-microfluidic platforms enable the high-resolution isolation and analysis of biological samples in nanoliter to picoliter volumes, allowing for the study of individual cells in controlled microenvironments and high-throughput screening. Notably applications, microwell platforms have been used for comprehensive single-cell analyses, including genomics, transcriptomics, proteomics, metabolomics, phenomics and cell–cell interactions, thereby enabling the exploration of cellular heterogeneity and providing deeper insight into cell population, complex biological tissue and disease heterogeneity[13–22].

For example, in single-cell transcriptomics, DeKosky et al.[23] employed a sealed microwell system to capture mRNA from individual B cells with magnetic poly(dT) beads. Emulsion PCR was then used to link and sequence heavy and light chain pairs, allowing single-cell B cell receptor (BCR) identification. Moreover, Gierahn et al.[24] addressed the limitations of oil-based sealing by developing Seq-Well, a microwell platform sealed with a semi-permeable membrane, which is easy to implement and enhances transcript capture and portability. In single-cell proteomics, for example, Love et al.[25] developed a strategy for identifying hybridoma cells secreting antigen-specific antibodies by covering microwell arrays with functionalized slides. Upon removal and imaging of the slide, secretion signatures were matched to individual microwell, allowing retrieval of live, target-specific cells for downstream clonal expansion.

Other notable applications, microwells have frequently been employed to isolate single microbeads for calibration purposes and to support biochemical assays [26–28]. For instance, Chung et al.[29] achieved over 80% cell trapping efficiency using 15 µm beads, while Park et al.[30] reported 62% single-cell capture with 10 µm beads across varying geometries. Fluorescent beads have also enabled visualization and calibration of loading efficiency[31]. Beyond trapping, microbeads improve molecular assays through high surface-area and customizable chemistries[32]. For instance, Shao et al.[33] used antibody-coated beads to detect six cytokines from single T cells, and Zhao et al.[34] integrated DNA-barcoded bead arrays into PDMS microwells to quantify ten proteins from individual macrophages.

The successful implementation of these studies depends critically on the CAD design of microfluidic microwell devices. However, being slow, error-prone, and dependent on specialized expertise, manual CAD workflows for microfluidic microwell devices make design modifications tedious and poorly reproducible, creating significant barriers to throughput, collaboration, and innovation. Furthermore, accurate classification of microwells by single cell or microbead occupancy, whether single, multiple, or empty, becomes essential; while single-bead wells enable discrete molecular analyses with minimal signal ambiguity, multiple-bead wells introduce risks of cross-reactivity, barcode collisions, and compromised quantification. Empty wells, though less disruptive, lower assay efficiency, and may indicate loading or design flaws. Accurate classification of these occupancy states is crucial for ensuring quality control and optimizing assays in high-throughput and single-cell workflows. A variety of methods have been developed to distinguish empty microwells from those containing single or multiple particles/cells. These approaches include intensity-based thresholding[35], Poisson-based statistical counting[35], template matching[36], correlation-based detection[36], and deep learning algorithms[37,38]. Since these classification methods rely on thresholding, handcrafted features, or supervised models trained on narrow image sets, they are highly sensitive to imaging variability and dependent on extensive manual annotation. Such annotation introduces bias and curtails scalability, while the technical skills needed for current machine learning workflows further constrain iterative experimentation and wider adoption.

By training on massive corpora, large language models, predominantly transformer-based deep learning architectures[39], learn statistical patterns of language, capturing syntax, semantics, and even some reasoning capabilities. Upon fine-tuning or through prompt engineering, LLMs can perform a diverse range of tasks, including text generation, summarization, translation, question answering, code generation, and reasoning. LLMs are rapidly reshaping a broad range of fields, including biomedical research[40–43], mechanical design[44,45], material science[46–52], chemistry[53–57], biology[58–62] and medicine[63–68] by enabling new avenues for scientific discovery and technological advancement[69–72]. LLM-based agents are autonomous or semi-autonomous systems that leverage LLM capabilities to plan, decide, and execute tasks by interacting with external tools, APIs, or environments. These agents have been applied to medical research[73–75], and autonomous scientific discovery[76]. Transitioning from language to vision-language reasoning, multimodal LLMs (MLLMs)[77–83] models capable of interpreting text, images, and other modalities, have further extended the frontier of machine intelligence. For instance, MLLMs have been applied to enhance image classification accuracy[84–86], medical image interpretation[87–89], document analysis[90–92], and the development of multimodal agents[93–97]. Despite their growing impact in other domains, the application of LLM agents and MLLMs into microwell microfluidic systems remains limited, highlighting a compelling opportunity for technological advancement.

Comparing traditional automated design frameworks for the automation and validation of microfluidic devices reveals that they are typically implemented as closed, rule-based systems[98–100], LLM Agents are capable of integrating complex tasks into autonomous, closed-loop digital discovery platforms such as literature synthesis, hypothesis generation, autonomous design, execution in self-driving laboratories[101–110], analysis of results, and the generation of new hypotheses.

In this study, we propose an LLM-driven framework for microwell design (Fig. 1), which integrates a mixture of LLM agents to generate CAD scripts for microwell geometries. Using this framework, we designed microwells of diverse shapes and dimensions prior to fabrication and subsequently loaded them with microbeads. Only through the use of state-of-the-art MLLMs were we able to classify microwell images directly, distinguishing between occupancy states (empty, single-bead, or multiple-bead) as well as their geometrical configurations. By integrating MLLMs with logistic regression, we designed a simplified classification approach that, in its final form, delivered marked improvements in accuracy. This study highlights the broader potential of LLMs and MLLMs in offering autonomous, scalable, cost-effective, and interpretable solutions for device prototyping and biomedical image analysis. This LLM Agents-driven framework enables a more adaptive and scalable framework, paving the way for accelerated innovation in microfluidic research.

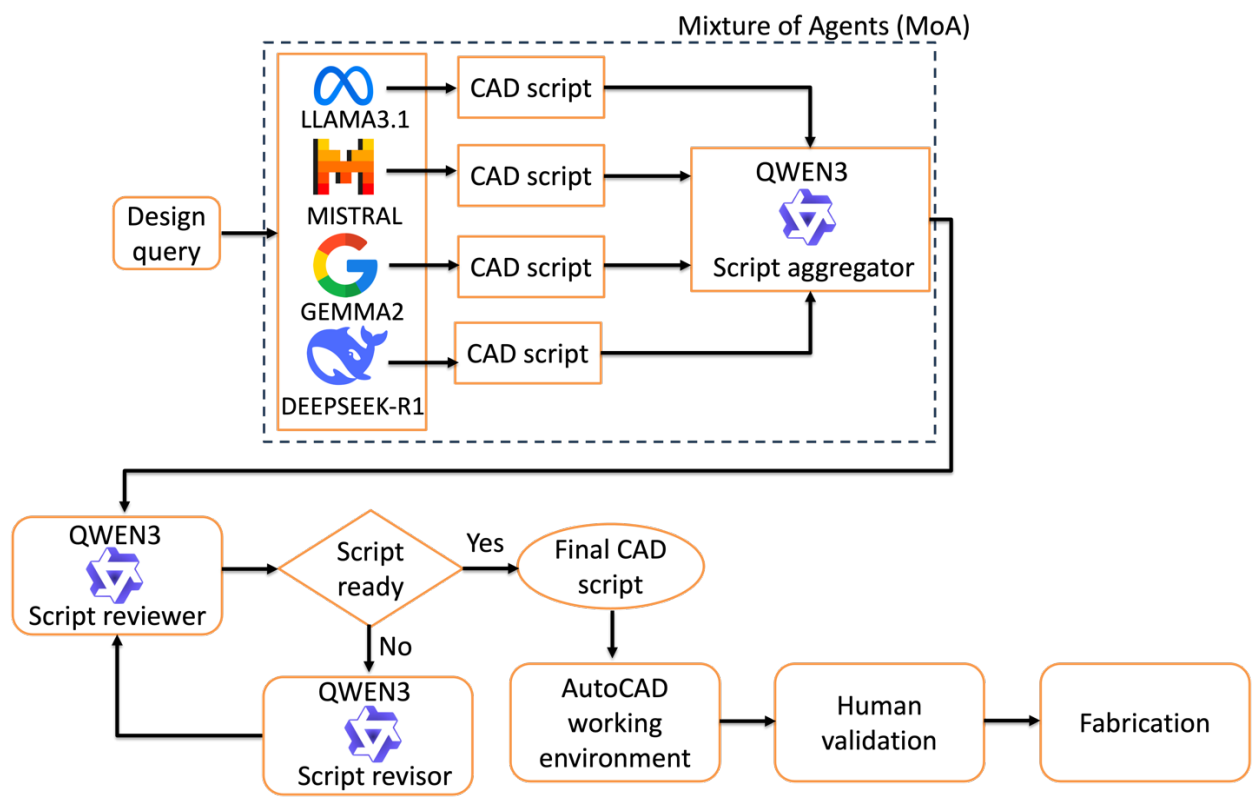

Fig. 1 LLM-driven microwell design framework

## 2. Methodology

### 2.1 Models used

In prioritizing rigorous scientific validation, the study employed only open-source models, five foundational LLMs for CAD design and three foundational MLLMs for classification, ensuring transparency, full access to architectures and weights, and reproducibility. In avoiding commercial

API costs, the approach achieves scalability and affordability, thereby facilitating long-term experimentation and community-wide access while adhering to principles of openness and reproducibility. A detailed list of the LLMs and MLLMs used is provided in Table 1.

| Task | Model | Model size | Vendor | License | Source |
|---|---|---|---|---|---|
| CAD design | GEMMA2 | 9B | Google | Open | [83] |
| | LLAMA3.1 | 8B | Meta | Open | [111] |
| | MISTRAL | 7B | Mistral AI | Open | [112] |
| | DEEPSEEK-R1 | 8B | DeepSeek | Open | [113] |
| | QWEN3 | 14B | Alibaba | Open | [114] |
| Classification | LLAMA3.2-vision | 11B | Meta | Open | [111,115] |
| | GEMMA3 | 4B | Google | Open | [116] |
| | LLAVA | 7B | Microsoft | Open | [79] |

Table 1. LLMs and MLLMs used in the experiment.

## 2.2 LLM-driven microwell design framework

The framework represents a sophisticated and forward-looking pipeline for generating AutoLISP scripts from natural language design queries, integrating multiple stages of LLM collaboration, automated review, and real-world validation as illustrated in Fig.1. At the heart of this architecture is a mixture of agents strategy[117], comprising LLAMA3, MISTRAL, and GEMMA2, which captures a diverse spectrum of model capabilities and linguistic priors. This ensemble strategy introduces diversity in reasoning patterns, reduces model-specific biases, and provides fault-tolerant redundancy, which is particularly critical for high-stakes domains such as design automation, where a single incorrect output may propagate structural or functional errors downstream. Not as a mere consolidator, but as a semantic synthesizer, does the LLAMA3.1-driven aggregation layer operate, unifying conflicting or partial fragments into a coherent and executable script. As it entails critical evaluation across outputs, this stage reflects meta-reasoning capabilities and, in doing so, lays the groundwork for ensemble learning in language-to-code translation. While consolidation is important, the key feature of the framework is a recursive feedback mechanism that continually refines the aggregated script through LLAMA3.1 until it reaches predefined quality standards, thereby serving as an automated peer review process. While many LLM-based pipelines stop at generation, this framework extends into structured validation, a critical requirement for deployment in engineering contexts. The final script undergoes a multi-layered verification process: it is first executed in a live AutoCAD environment to confirm syntactic and functional validity, and only thereafter is it reviewed by human experts to ensure adherence to domain-specific constraints, design intent, and safety before fabrication. These layers form a hybrid validation scheme that blends machine execution with expert oversight, addressing key limitations of purely LLM-based systems. A detailed explanation of the LLM-driven microwell design framework's prompts is provided in the Table S1 in the supplementary file.

**2.3 Direct inference with MLLMs for classification of microwell occupancy and shape**

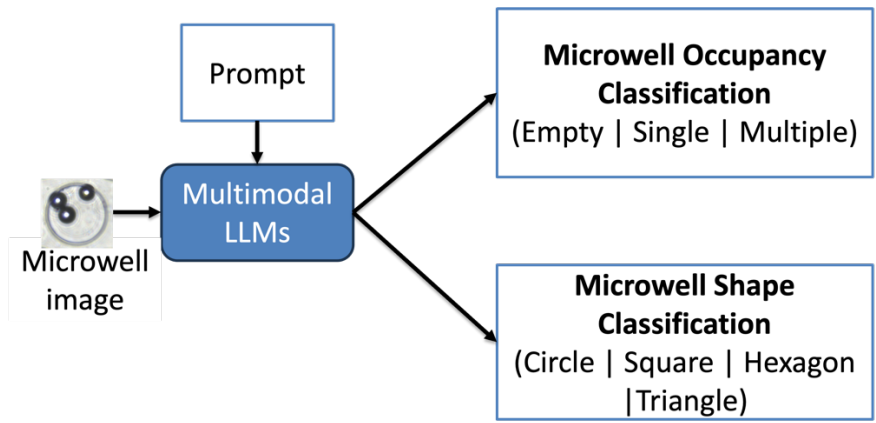

(a)

```
prompt = (
    "Classify this microwell based on the shape and the count of bright microbeads. "
    "Respond only with JSON: {\"count\":<empty|single|multiple>,"
    "\"shape\":<circle|square|triangle|hexagon>}."
)
```

(b)

Fig. 2 (a) Diagram of direct inference with MLLMs for classification of microwell occupancy and shape. (b) Prompt used

The MLLM-based framework and its used prompt for automated microwell image classification are illustrated in Fig. 2(a) and Fig. 2(b), respectively. The system integrates microwell images and tailored prompts as inputs to the MLLMs, enabling two distinct classification tasks: (1) microwell count classification, categorizing images into empty, single, or multiple microwell conditions; and (2) microwell shape classification, determining the geometric form of microwells as circle, square, hexagon, or triangle. The schematic illustrates the prompt-conditioning mechanism that guides the MLLM towards the intended visual recognition task, demonstrating the flexibility and generalization capacity of vision-language models for specialized biomedical imaging analyses.

**2.4 MLLM-logistic regression classification framework**

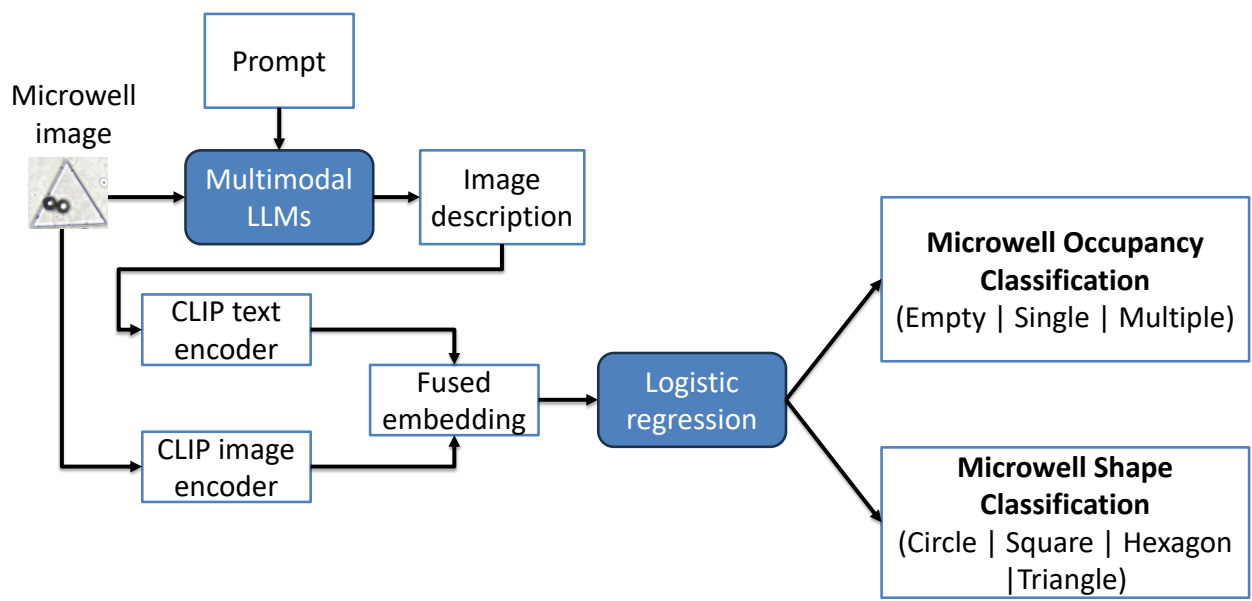

(a)

```
prompt = (
"Instruction: Specify whether the microwell is square,
circular, hexagonal, or triangular. Describe the main
visible object in the microwell image with precise and
specific language. "
"Respond using no more than two concise sentences.
Avoid unnecessary elaboration."
)
```

(b)

Fig. 3 (a) MLLM-logistic regression classification framework for classification of microwell occupancy and shape. (b) Prompt used

A comprehensive multimodal framework and a designed prompt, developed to advance microwell image classification by effectively integrating the semantic reasoning capabilities of MLLMs with an efficient and interpretable machine learning classifier, are shown in Fig. 3(a) and Fig. 3(b), respectively. By first providing a microwell image together with a strategically constructed prompt to an MLLM, the workflow yields a semantic description that conveys higher-level contextual, morphological, and geometric details, such as counts, arrangements, and precise shape characteristics that conventional vision-based models frequently fail to capture. Through a CLIP-based text encoder[118], the generated description is converted into semantic embeddings, whereas the corresponding microwell image is simultaneously processed by a CLIP image encoder[118] to capture fine-grained spatial and structural representations. After encoding both modalities, they are fused to form a joint multimodal embedding that integrates semantic insights derived from language with detailed pixel-level visual information. By utilising descriptions generated from MLLMs, the framework achieves a key advantage, as it integrates advanced visual reasoning that surpasses the capabilities of purely vision-based approaches. Moreover, fusion with direct image embeddings ensures the preservation of critical fine-grained spatial features necessary for robust classification performance. The fused multimodal embedding is then fed into a logistic regression model[119,120] to perform two essential classification tasks: microwell count classification, categorizing wells as empty, containing single objects, or containing multiple objects, and microwell shape classification, identifying wells as circles, squares, hexagons, or triangles. Another significant strength of this framework is that, by deploying logistic regression for the final classification step, the framework gains a highly intuitive and interpretable model that requires minimal computational resources and allows for rapid training and inference. This design choice substantially reduces hardware requirements and energy consumption while maintaining competitive classification accuracy, due to the rich and informative fused embeddings provided by the upstream multimodal feature extraction. Overall, this framework exemplifies an innovative, scalable, and resource-efficient solution for high-throughput microwell image analysis, with broader implications for biomedical imaging pipelines. Its integration of powerful semantic-contextual reasoning with direct visual feature representation, coupled with an interpretable and computationally economical classifier, provides a generalisable approach for automated laboratory assays and vision-language analysis tasks where performance, interpretability, and resource efficiency are simultaneously critical.

**2.5 Image dataset preparation for classification**

We employed our LLM-driven microwell design framework to generate geometries of varying shapes, including circles, squares, triangles, and hexagons with circumradii ranging from 10 to 35 μm in defined increments. SU-8 molds with a pattern height of 20 μm were fabricated based on these designs and subsequently used to cast PDMS microwell chips. These chips were then loaded with $5 \times 10^6$ microbeads/mL (15 μm in diameter). After a 20-minute settling period, microwell images were captured across all shape and size variants. A total of 2,200 images, equally distributed across the four shapes and three occupancy states (empty, single microbead, multiple microbeads), were extracted, curated, and augmented to form the dataset. This dataset was used for direct inference using MLLMs. For the proposed MLLM-logistic regression classification framework, the data were split into 80% for training and 20% for testing. Repeated 10-fold cross-validation was performed 15 times to ensure robust and reliable performance estimates by mitigating variability introduced by random sampling[120].

**3. Results and discussion**

**3.1 CAD script generation**

| Design query | | | |
|---|---|---|---|
| Generates an array of **400 square** arranged in a **rectangular lattice pattern**. Each square has a **side length of 25 micrometers**. The **center-to-center distance between adjacent squares** is **80 micrometers**. The array should fill as many rows and columns as necessary to reach exactly 400 squares. | | | |
| **Responses by LLMs** | | | |
| GEMMA2 | MISTRAL | LLAMA3.1 | DEEPSEEK-R1 |
| (defun c:MicrowellArrayDesign () ;; Define design parameters (setq side_length 25) ; Side length of each square in micrometers (setq spacing 80) ; Center-to-center distance between squares in micrometers (setq total_squares 400) ; Total number of squares ;; Calculate the number of rows and columns (setq num_rows (sqrt total_squares)) (setq num_cols (ceiling (/ total_squares num_rows))) ;; | (defun c:MicrowellArrayDesign () ;; Define fixed parameters (setq sideLength 25 ;; Side length of each square in micrometers spacing 80 ;; Center-to-center distance between adjacent squares in micrometers totalSquares 400 ;; Total number of squares rowCount 0 ;; Initialize row counter columnCount 0 ;; Initialize column counter currentX 0 ;; Initialize current X position currentY 0 ;; Initialize current Y position );; | (defun c:MicrowellArrayDesign () (setq side-length 25 ; Side length of each square in micrometers spacing 80 ; Center-to-center distance between adjacent squares in micrometers total-squares 400) ; Total number of squares to be generated ; Calculate the number of rows and columns required to reach the total number of squares (setq num-rows (sqrt (/ total-squares (floor (/ total-squares (floor (/ total-squares 1)))))) | defun c:MicrowellArrayDesign () ; Set design parameters setq *side-length* 25 setq *spacing* 80 setq *total-squares* 400 ; Calculate number of rows and columns needed setq *rows* (ceiling (sqrt *total-squares*)) setq *columns* (ceiling (* *rows* *rows*)) ; Generate the array using nested repeat loops repeat for *row* 0 to *rows*-1 repeat for *col* 0 to *columns*-1 ; Calculate starting position for each square |

| QWEN3-Aggregator | QWEN3-Reviewer | QWEN3-Revisor |
|---|---|---|
| (defun c:drawarray () ;; Define parameters (setq total 400) ; Total number of squares (setq spacing 80) ; Center-to-center spacing (micrometers) (setq side 25) ; Side length of each square (micrometers) ;; Calculate number of rows and columns (setq rows (fix (sqrt total))) ; Number of rows (setq cols (/ total rows)) ; Number of columns ;; Initialize row counter (setq i 0) ;; Outer loop for rows (repeat rows ;; Inner loop for columns (setq j 0) (repeat cols ;; Calculate center coordinates for current square (setq x (* j spacing)) ; X-coordinate of center (setq y (* i spacing)) ; Y-coordinate of center | #### 1. **Function Definition** - The function is correctly defined as `(defun c:MicrowellArrayDesign () ...)` with a valid command name (`c:MicrowellArrayDesign`). This allows it to be executed as a command in AutoCAD. #### 2. **Syntax Validity** - All AutoLISP syntax is valid: - `setq` is used correctly for variable assignments. - `repeat` is used for loops. - `command` is used to invoke AutoCAD commands. - `fix` is used for integer truncation, which is valid in AutoLISP. | (defun c:MicrowellArrayDesign () ;; Define parameters for the microwell array (setq total 400) ; Total number of squares in the array (setq spacing 80) ; Center-to-center spacing between squares (micrometers) (setq side 25) ; Side length of each square (micrometers) ;; Calculate number of rows and columns ;; Rows are calculated as the integer square root of total (setq rows (fix (sqrt total))) ;; Columns are calculated as total divided by rows (setq cols (/ total rows)) ;; Initialize row counter (setq i 0) ;; Outer loop for rows (repeat rows ;; Initialize column counter for each row |

Fig. 4 Workflow for creating an AutoLISP script

An example of synthesizing an AutoLISP script that generates a hexagonal array of circles is illustrated in Fig. 4. The process begins with a design query specifying the objective of writing a script to implement the desired geometric array. In the subsequent stage, multiple LLMs (LLAMA3.1, MISTRAL, GEMMA2, DEEPSEEK-R1) independently produce candidate solutions in response to the design query. While operating in parallel, these models each offer a unique interpretation and implementation strategy, thereby exploiting architectural and training diversity to generate complementary perspectives on the problem. The candidate scripts generated by each model are then processed by the QWEN3-Aggregator, a semantic synthesiser tasked with integrating these disparate outputs. The Aggregator performs reconciliation of conflicting implementations and resolution of incomplete fragments, synthesising them into a single, unified, coherent, and executable AutoLISP script. The synthesised script undergoes a review phase by the QWEN3-Reviewer, which evaluates the script for correctness, completeness, logical consistency, and adherence to the original design specification. While the Reviewer detects errors, inefficiencies, or ambiguities that may compromise execution or deviate from the intended geometry, the QWEN3-Revisor applies the suggested corrections in the final stage. This involves refining the code structure, optimizing syntactic and semantic clarity, and making final adjustments to ensure the script is fully executable within the AutoLISP environment, robust to edge cases, and compliant with geometric precision requirements. An example of the script-generated CAD being used to fabricate SU-8 wafer molds and PDMS microwells is shown in Fig. 5. These microwells were subsequently loaded with microbeads, and the resulting microwell–microbead images were then employed for classification analyses.

Script-generated CAD 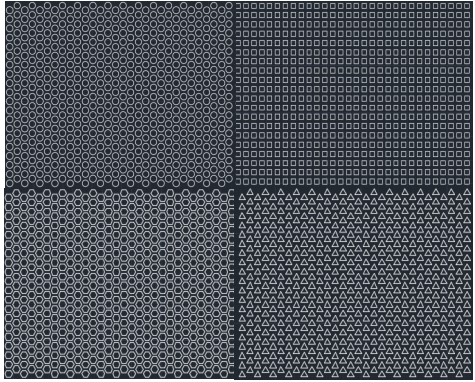

SU-8 wafer mold 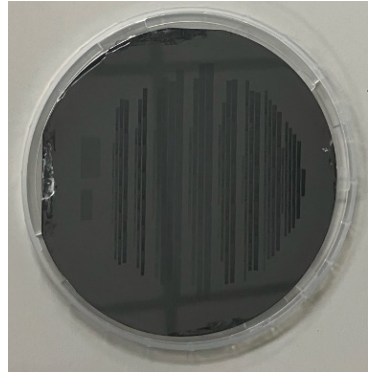

Microwell image 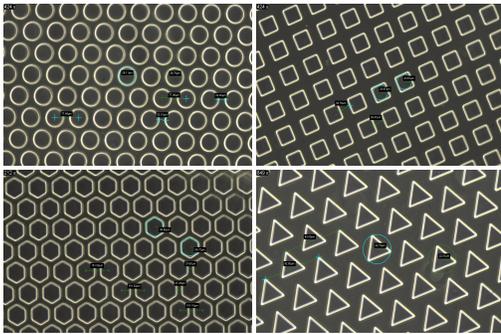

PDMS microwell 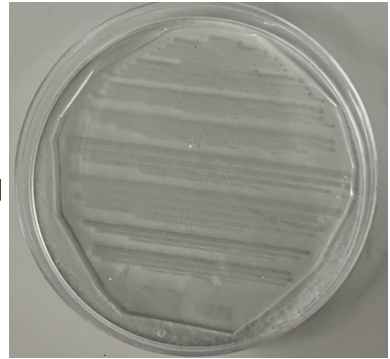

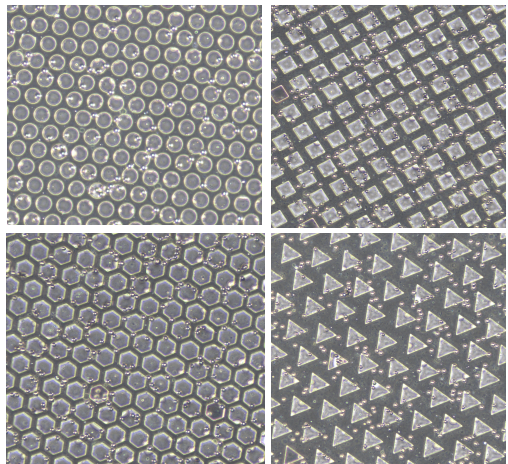

Microwell image after microbead loading

Fig. 5 From script to experiment: CAD-generated SU-8 molds and PDMS microwells for microbead loading.

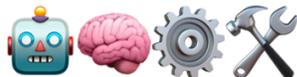

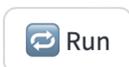

Fig. 6 Graphical user interface (GUI) of autonomous microwell design agent

We implemented a minimal yet intuitive graphical user interface (GUI) to support straightforward user interaction (Fig. 16). The associated GitHub repository includes comprehensive installation instructions in the README documentation and a demonstration video illustrating GUI usage.

**3.2 Accuracy of direct inference with MLLMs for classification of microwell occupancy and shape**

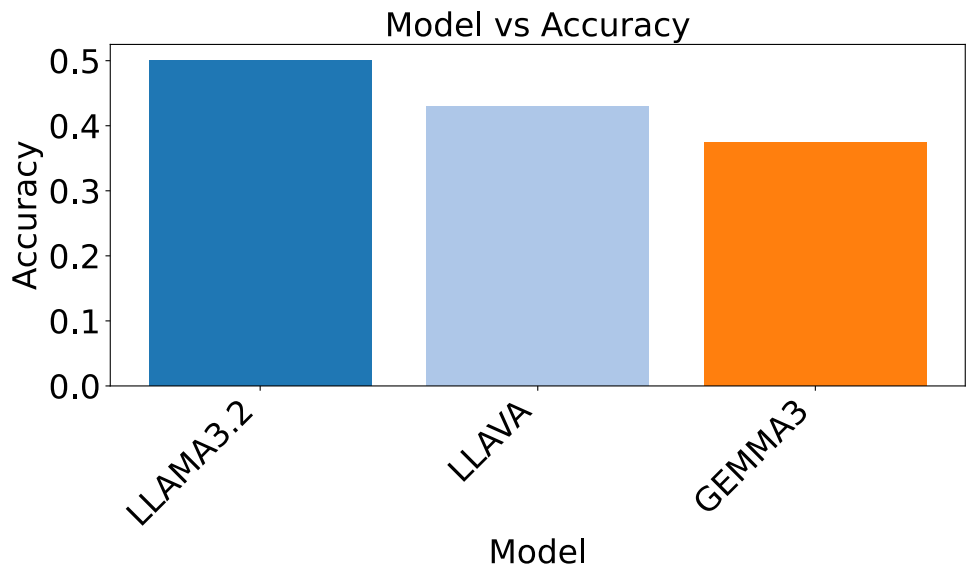

Fig. 7 Accuracy for microwell occupancy classification

A comparative evaluation of the microwell occupancy classification accuracy across three cutting-edge MLLMs: LLAMA3.2-vision:11B, LLAVA 7B**,** and GEMMA3:4B as shown in Fig. 7**.** The task involves classifying microwell images into *empty, single,* or *multiple* occupancy categories. LLAMA3.2-vision:11B achieved the highest accuracy at approximately 0.50, followed by LLAVA:7B at around 0.43, and GEMMA3:4B with the lowest performance at around 0.37. While achieving the top position among evaluated models, LLAMA3.2-vision nonetheless exhibits accuracy far below deployment thresholds, highlighting the limitations of general-purpose multimodal LLMs in microscopy-based object quantification. LLAVA:7B, a vision-language aligned variant trained on image-question answering datasets, underperforms LLAMA3.2-vision:11B despite being designed for visual reasoning, raising concerns about its generalization capacity to domain-specific imaging modalities, such as microscopy. GEMMA3:4B's poor performance suggests that its current multimodal integration capabilities are insufficient for object enumeration tasks in dense or low-contrast images typical of microwell assays. While multimodal LLMs demonstrate promising general visual understanding, they remain inadequate for rigorous quantitative bioimaging analysis without domain-adaptive fine-tuning, explicit architectural augmentations for spatial reasoning, or integration with traditional computer vision pipelines. Additionally, the limited performance across models expresses that simply increasing model size or vision-language alignment parameters does not suffice for specialized biomedical applications, underscoring the need for training on microscopy-specific datasets with task-focused objectives.

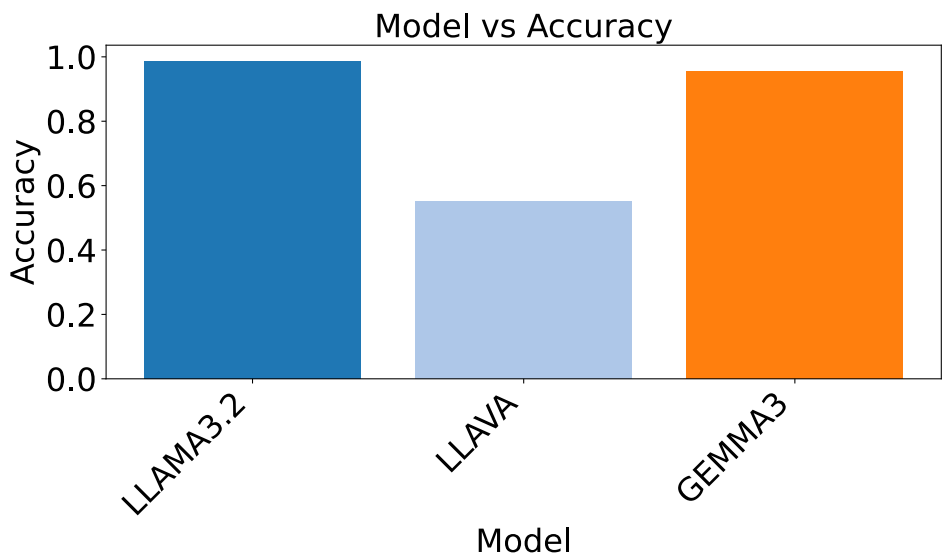

Fig. 8 Accuracy for microwell shape classification

With accuracy results illustrated in Fig. 8**,** LLAMA3.2-vision:11B is shown to surpass GEMMA3:4B and LLAVA:7B, achieving nearly 0.99 accuracy and demonstrating its robust capability to extract fine-grained morphological features critical for microwell shape discrimination. While GEMMA3:4B demonstrated strong performance at nearly 0.95**,** LLAVA:7B achieved only around 0.55, exposing substantial limitations in handling microscopy data even with its larger parameter scale. Although larger scale may appear advantageous**,** this discrepancy demonstrates that architecture, corpus relevance, and vision–language alignment ultimately govern downstream accuracy. As these results reveal**,** applying general-purpose MLLMs to high-resolution biomedical imaging requires careful model selection, domain-relevant pretraining, and fine-tuning.

## 3.3 Enhanced classification of microwell occupancy and shape using the MLLM-logistic regression classification framework

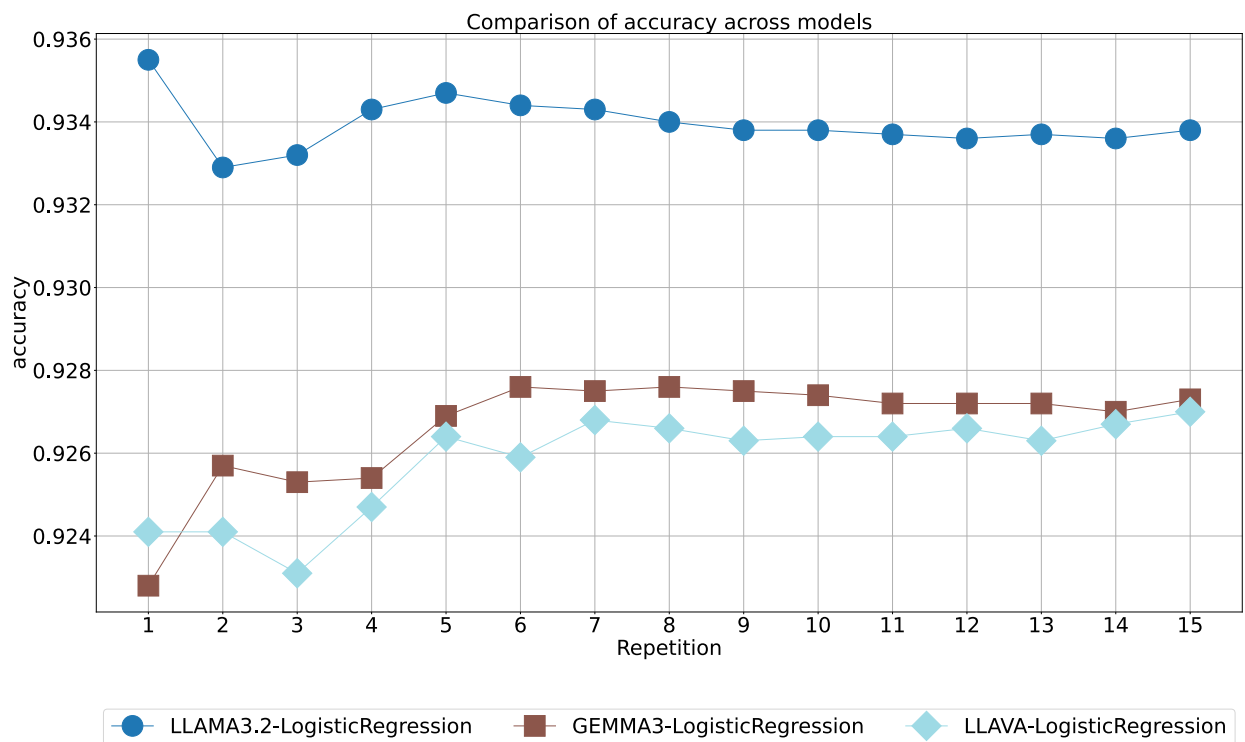

Fig. 9 Robust comparison of model accuracy for microwell occupancy classification using repeated 10-fold cross-validation.

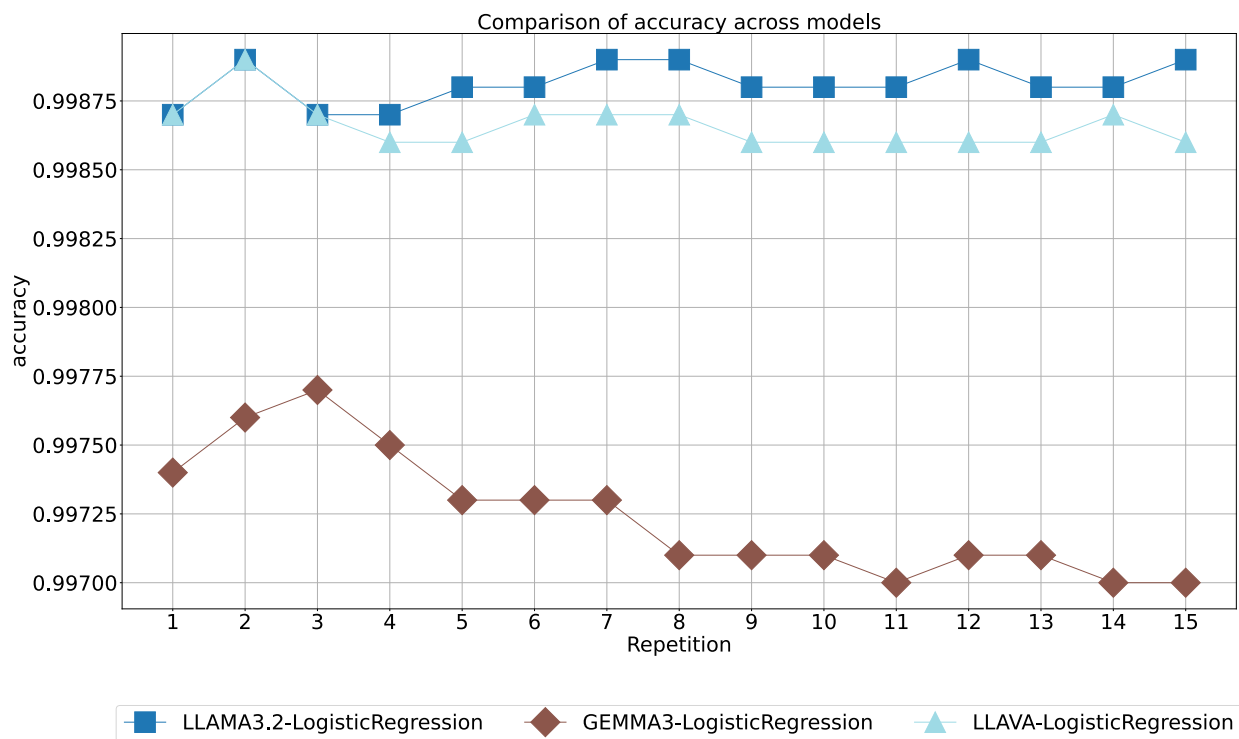

Fig. 10 Robust comparison of model accuracy for microwell shape classification using repeated 10-fold cross-validation.

A detailed comparison of classification accuracy across 15 repetitions of repeated 10-fold cross-validation for three MLLM-based pipelines applied to the task of microwell count classification is illustrated in Fig. 9. Each model in the framework is tasked with generating textual descriptions of microwell images via prompting, which are then encoded using a CLIP text encoder. Simultaneously, the input image is processed through the medium of a CLIP image encoder. These two modalities are fused into a joint representation and passed to a logistic regression classifier for downstream classification. Notably, all models share identical image and text encoders and classifier architecture, differing only in the LLM responsible for generating the image captions, thereby isolating the impact of textual quality on classification performance. With accuracies stabilising between 0.934 and 0.935, LLAMA3.2-vision:11B exhibits negligible variance and clearly outperforms GEMMA3:4B and LLAVA:7B. With accuracies exceeding 0.9987 and minimal variance, LLAMA3.2-vision:11B again dominates in microwell shape classification (Fig. 10), its consistent advantage demonstrating the critical influence of textual description quality in multimodal fusion pipelines. LLAMA3.2-vision:11B likely generates more precise, detailed, and semantically aligned descriptions of microwell images, capturing nuanced cues such as cell morphology, quantity, and spatial arrangement, which leads to more discriminative text embeddings that enhance alignment with the image representation. A further and critical strength of this framework is its use of a *simple, intuitive logistic regression classifier*. Despite its linear nature and low model complexity, logistic regression performs remarkably well in this setup. This outcome does reveal that the fused multimodal embeddings, when built upon high-quality, semantically grounded textual descriptions, are sufficiently linearly separable to allow a minimalist model to succeed. This has important implications. First, it reduces the risk of overfitting, which is often associated with deep neural network-based classifiers. Second, it enhances transparency and interpretability, which are essential for biomedical applications where model decisions must be auditable and clinically justifiable.

## 4. Conclusion and Outlook

Since the use of LLMs and MLLMs in microwell microfluidics has not been systematically explored, despite their broad applicability, we developed a framework that utilizes LLMs to automate CAD script generation, enabling rapid microwell fabrication with varied geometries, followed by microbead loading. When descriptive outputs from MLLMs were combined with image embeddings and analysed by logistic regression, a method selected for its efficiency and interpretability, accuracies rose dramatically to greater than 0.92 for occupancy and 0.99 for shape, in contrast to the modest values (0.5 and 0.55) obtained with direct classification. Not only do LLMs streamline CAD workflows in microwell microfluidics, but, as these findings demonstrate, MLLMs also provide an overlooked capacity to extract complex biological data from imaging, jointly driving new avenues in discovery and computation.

Integrating LLM agents into CAD workflows changes how microfluidic systems, and the microwell platform in particular, are conceptualized and developed. Traditionally, the design of microfluidic devices has required significant domain expertise, as well as proficiency in CAD software and parametric scripting languages. To address these challenges, LLMs, refined on massive collections of computational, scientific, and engineering materials, can be employed for converting commands into functional and valid CAD scripts[44,45,121–123]. This enables the rapid prototyping of diverse geometries of well arrays for optimization for domain-specific biological and analytical purposes. For the capabilities of LLM agents, not only can they generate code, but by repeatedly refining design parameters based on user feedback or performance data, they also establish a closed-loop optimization cycle that improves reproducibility, accelerates development, and enhances traceability. These essential attributes can advance translational research while ensuring regulatory compliance. Additionally, LLMs can be smoothly adapted to specific domains through proper fine-tuning techniques and integration of tools. Doing so enables LLMs to become intelligent co-designers in the field of microfluidic device engineering, allowing for unprecedented bridging of computational design with experimental fabrication.

The capability to process both texts and images enables MLLMs to accomplish a higher-level semantic understanding of image content. This results in multiple advantages in biomedical image analysis[87–89,124–127]. First, the variability and subjectivity inherent to manual annotation or conventional computer vision approaches can be reduced. Second, the outputs of microwell assays, such as temporal dynamics, fluorescence signals, and heterogeneous cell distributions, can be interpreted. Additionally, occupancy state identification, classification of cell morphologies, and generation of diagnostic information or descriptive summaries can be accomplished with minimal human oversight. These potentials of MLLMs could facilitate unified precision medicine pipelines[128–130]. Together with the optimization of integrated precision medicine pipelines, as the autonomy in lab-on-a-chip technologies is significantly gained, the deployment of MLLMs can support real-time decision-making, adaptive experimentation, and ultimately, intelligent bioassay platforms.

**Conflicts of interest**

There are no conflicts to declare.

**Data availability**

The datasets and code for the analyses and figure generations in this work are publicly available on GitHub at url: https://github.com/duydinhlab/MicrowellAgent

**Corresponding Author:** ngocduydinh@cuhk.edu.hk

Address: BME office, Room 1120, 11/F, William M.W. Mong Engineering Building, or Room 208, Ho Sin Hang Engineering Building (SHB), The Chinese University of Hong Kong, Shatin, N.T., Hong Kong

**Author Contributions:**

Conceptualization – N-D.D.; Methodology – D-N.N., N-D.D.; Investigation – D-N.N.; Data curation – D-N.N.; Writing – original draft – D-N.N.; Writing – review & editing – D-N.N., N-D.D.; Supervision – S.S., R.K-Y.T., N-D.D; Funding acquisition – R.K-Y.T., N-D.D.

**Acknowledgements**

We gratefully acknowledge the funding provided by the Research Grant Council of Hong Kong, General Research Fund (Ref No. 14211223).

**Table of Content**

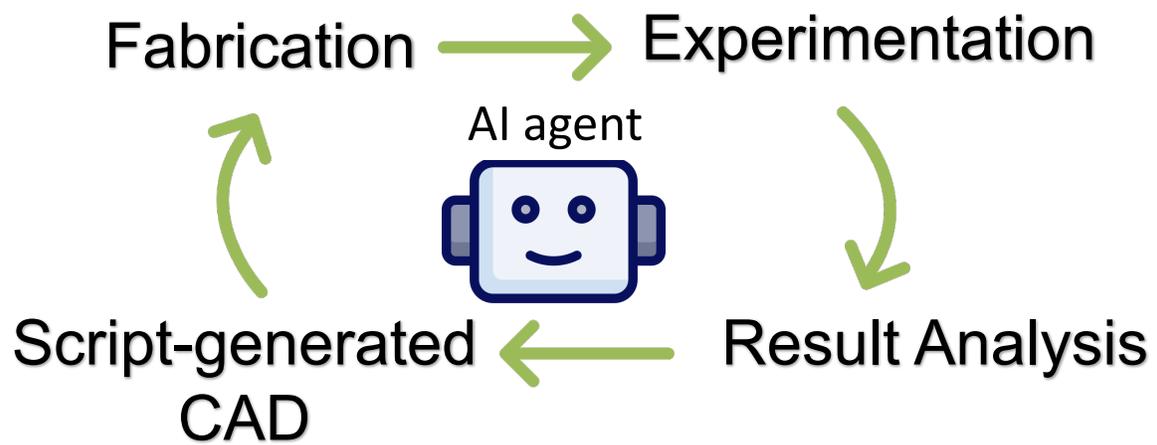

# Large Language Model Agents Enable Autonomous Design and Image Analysis of Microwell Microfluidics

Dinh-Nguyen Nguyen[1], Sadia Shakil[1], Raymond Kai-Yu Tong[1*], Ngoc-Duy Dinh[1*]

[1]Department of Biomedical Engineering, The Chinese University of Hong Kong, Shatin, N.T., Hong Kong, China

*Corresponding Author

Corresponding Email: ngocduydinh@cuhk.edu.hk

Table S1. Prompts of LLM-driven microwell design framework

| Components | Prompt |
|---|---|
| LLAMA3, MISTRAL, GEMMA2, DEEPSEEK-R1 | You are an expert AutoLISP programmer. Your task is to generate AutoLISP code that can be copied directly into the Visual LISP Editor in AutoCAD and run without errors.<br><br>Rules:<br><br>1.The script must be defined as a function using 'defun c:MicrowellArrayDesign ()' so it can be executed as a command in AutoCAD.<br><br>2.Do not include /localvars in the function definition.<br><br>3.All design parameters (e.g., side length, spacing, number of rows, number of columns, or total number of objects) must be set inside the function using setq. Do not request user input.<br><br>4.Use only valid AutoLISP syntax: |
| | 5. Use setq, repeat, if, while, command, entmake.

6. Do not use Common Lisp–only constructs such as lambda, loop, error, ceiling, or advanced let scoping.

7. All geometry must be created with AutoCAD commands (command) or entity creation functions (entmake) using proper DXF codes.

8. The code must run without syntax errors in the Visual LISP Editor. Always balance parentheses.

9. Add clear comments to explain each section of the code.

Example task format:

Define fixed parameters with setq.

Use repeat loops to generate rows and columns.

Place geometry (e.g., RECTANGLE or SOLID) at correct positions based on spacing.

Ensure the final array matches the specification (e.g., exactly 400 squares).

Deliverable:

Provide the complete AutoLISP function, ready to paste and run in AutoCAD.

Question:

{question} |
| QWEN3 aggregator | You are a You are an expert AutoLISP programmer and reviewer. You have received multiple responses generated by different models for the same question.

Question:

{question}

Response 1:

{response_1}

Response 2: |

{response_2}

Response 3:

{response_3}

Response 4:

{response_4}

Your task is to synthesize a single high-quality answer that meets the following requirements:

<GOAL>

-Combine the most accurate and relevant parts of the responses into one refined answer.

-The script must be defined as a function using 'defun c:MicrowellArrayDesign ()' so it can be executed as a command in AutoCAD.

-Do not include /localvars in the function definition.

-All design parameters (e.g., side length, spacing, number of rows, number of columns, or total number of objects) must be set inside the function using setq. Do not request user input.

-Use only valid AutoLISP syntax:

-Use setq, repeat, if, while, command, entmake.

-Do not use Common Lisp–only constructs such as lambda, loop, error, ceiling, or advanced let scoping.

-All geometry must be created with AutoCAD commands (command) or entity creation functions (entmake) using proper DXF codes.

-The code must run without syntax errors in the Visual LISP Editor. Always balance parentheses.

-Add clear comments to explain each section of the code.

-AutoLISP code can be copied directly into the Visual LISP Editor in AutoCAD and run without errors.

| | |
|---|---|
| | \<GOAL\>

\<REQUIREMENTS\>

1. Evaluate the correctness and quality of each response critically—some may contain errors or redundant code.

2. Do not copy the responses verbatim. Instead, recompose a clean and optimized solution.

3. AutoLISP code that can be copied directly into the Visual LISP Editor in AutoCAD and run without errors.

4. The final output must be logically sound, syntactically correct, and follow AutoLISP best practices.

\<REQUIREMENTS\> |
| QWEN3 reviewer | You are an expert AutoLISP programmer. Your task is to review AutoLISP code to determine whether it will run successfully in the Visual LISP Editor in AutoCAD.

Review Instructions:

1.Check if the script is defined correctly with 'defun c:MicrowellArrayDesign ()' so that it can be executed as a command in AutoCAD.

2.Verify that the code uses only valid AutoLISP syntax (e.g., setq, repeat, if, while, command, entmake).

3.Identify any use of Common Lisp–only constructs ('lambda', 'loop', 'error', advanced 'let' scoping, 'ceiling', etc.) and flag them as invalid in AutoLISP.

4.Confirm that all parentheses are balanced and the function body is syntactically correct.

5.Check that entity creation is valid:

For command, ensure correct arguments for AutoCAD commands like "RECTANGLE".

For entmake, confirm valid DXF group codes for the entity type.

6.Verify parameter handling:

If parameters are set inside the function, confirm 'setq' is used correctly. |

| | |
|---|---|
| | If input functions are used, confirm correct ones ('getint', 'getreal', 'getpoint') instead of 'getstring' for numeric/coordinate input.

7.Point out any logic errors (e.g., undefined loop counters, using integers vs lists, repeat misuse).

8.In your review, state clearly:

Yes: The code will run in Visual LISP Editor.

No: It will not run — then list the errors and suggest corrections.

Deliverable:

Provide a clear and precise review of whether the given AutoLISP code will run in the Visual LISP Editor of AutoCAD, highlighting syntax errors, invalid constructs, and compatibility issues.

Here is the code:

{code} |
| QWEN3 revisor | You are an expert AutoLISP programmer. Your task is to improve and correct AutoLISP code so that it can be copied directly into the Visual LISP Editor in AutoCAD and run without errors.

Improvement Instructions:

1.MUST ensure the function is defined properly as 'defun c:MicrowellArrayDesign ()' so it can be executed as an AutoCAD command.

2.Remove any invalid Common Lisp constructs ('lambda', 'loop', 'error', advanced 'let' scoping, 'ceiling', etc.) and replace them with valid AutoLISP syntax.

3.Ensure all parameters (e.g., side length, spacing, number of rows/columns) are set inside the function using setq, unless the task requires user input.

4.Replace any incorrect input functions with AutoLISP-valid ones ('getint', 'getreal', 'getpoint') instead of 'getstring' for numeric or coordinate input.

5.Correct the use of repeat loops by ensuring counters are properly defined and incremented with setq.

6. Unit must follow the requirement (e.g., micrometer) |

| | For geometry creation: |
| --- | --- |
| | If using 'command', ensure the arguments match AutoCAD's expectations (e.g., "RECTANGLE" requires two corner points). |
| | If using 'entmake', ensure correct DXF group codes are provided for the entity type. |
| | 6. Verify that all parentheses are balanced and the function body is syntactically correct. |
| | 7. Add clear comments explaining each section of the improved code. |
| | Deliverable: |
| | Return the improved and runnable AutoLISP script function as 'defun c:MicrowellArrayDesign ()' so it can be executed as an AutoCAD command, ready to paste into the Visual LISP Editor, along with a brief explanation of the corrections you applied. |
| | Guidelines: |
| | {guidelines} |
| | Question: |
| | {question} |
| | Original Code: |
| | {code} |

Table S1. Prompts of LLM-driven microwell design framework